\documentclass[a4paper,fleqn]{cas-dc}

\usepackage[numbers]{natbib}

\usepackage{algorithm}
\usepackage{algpseudocode}

\def\tsc#1{\csdef{#1}{\textsc{\lowercase{#1}}\xspace}}
\tsc{WGM}
\tsc{QE}
\begin{document}
\let\WriteBookmarks\relax
\def\floatpagepagefraction{1}
\def\textpagefraction{.001}

\shorttitle{}    

\shortauthors{X. Xie et~al.}  

\title [mode = title]{Optimization and validation of charge transport simulation for hybrid pixel detectors incorporating the repulsion effect}  
% \title [mode = title]{Charge transport simulation for hybrid pixel detectors incorporating the repulsion effect: implementation, parameterization, and validation}
% \title [mode = title]{Why repulsion matters for charge transport simulation in silicon sensors}




\author[a]{X.~Xie} [
    orcid=0000-0001-6473-7886,]
\cormark[1]
\cortext[1]{Corresponding author}
\ead{xiangyu.xie@psi.ch}
\author[a]{A.~Bergamaschi} \author[a]{M.~Br\"uckner} \author[a]{M.~Carulla} \author[a]{R.~Dinapoli} \author[a]{S.~Ebner} \author[a]{K.~Ferjaoui} \author[a]{A.~Francesca~Mazzoleni} \author[a]{J.~Franklin~Mulvey} \author[a]{V.~Gautam} \author[a]{D.~Greiffenberg} \author[a]{S.~Hasanaj} \author[a]{J.~Heymes} \author[a]{V.~Hinger} \author[a]{V.~Kedych} \author[a]{T.~King} \author[a]{S.~Li} \author[a]{C.~Lopez-Cuenca} \author[a]{D.~Mezza} \author[a]{K.~Moustakas} \author[a]{A.~Mozzanica} \author[a]{M.~M\"uller} \author[a]{K.A.~Paton} \author[a]{C.~Ruder} \author[a]{B.~Schmitt} \author[a]{P.~Sieberer} \author[a]{S.~Silletta} \author[a]{D.~Thattil} \author[a]{~J.~Zhang} \author[a]{E.~Fr\"ojdh}




\affiliation[a]{organization={Photon Science Detector Group, Paul Scherrer Insitute},
            addressline={Forschungsstrasse 111}, 
            city={Villigen},
            postcode={5232}, 
            country={Switzerland}}




\begin{abstract}
    For emerging applications of hybrid pixel detectors which require high spatial resolution, e.g., subpixel interpolation in X-ray imaging and deep learning-based electron localization, accurate modeling of charge transport processes in the sensor is highly demanded. 
    To address this, two open-source, time-stepping Monte Carlo simulation methods have been developed, both explicitly incorporating charge repulsion, which are found necessary for accurate simulation when charge sharing becomes important.
    The first method employs brute-force calculations accelerated by GPU computing to model charge carrier dynamics, including drift, diffusion, and repulsion.
    The second utilizes a simplified spherical model that significantly reduces computational complexity.
    A parameterization scheme of the charge transport behaviors has been developed to enable efficient and rapid generation of X-ray simulation events.
    Both methods were rigorously validated using experimental data collected with a monochromatic X-ray beam at the METROLOGIE beamline of the SOLEIL synchrotron, demonstrating excellent agreement with measured pixel-energy spectra across various sensor thicknesses, bias voltages, and photon energies.
    Furthermore, the impact of the repulsion effect on charge carrier distributions was quantitatively evaluated.
    The potential applications of these simulation methods for different particle detections and detector technologies are also discussed.
\end{abstract}


\begin{highlights}
\item Two open-source, time-stepping Monte Carlo simulation methods were developed to accurately model charge transport, explicitly incorporating repulsion, diffusion, and drift effects—one utilizing brute-force calculations accelerated by GPU computing, and the other employing a computationally efficient simplified spherical model.
\item Excellent agreement between simulated and measured pixel-energy spectra was obtained across various sensor thicknesses, bias voltages, and photon energies, validated with monochromatic X-ray beam data.
\item A parameterization scheme was developed, enabling rapid and efficient generation of accurate simulation events.
\end{highlights}

%\nocite{*}

% Keywords

\maketitle

% Main text

\section{Introduction}\label{}
Advanced applications of the MÖNCH detector \cite{MONCH}---a hybrid pixel detector with a 25 \textmu $\rm m$ pixel pitch and fast charge integrating readout---have demonstrated enhanced spatial resolution for X-rays through subpixel interpolation \cite{Chiriotti_2022} and for 200 keV electrons using deep learning techniques \cite{EM_deepLearning2023}.
In both cases, simulation events are of increasing importance, as they enable the generation of interpolation mappings and the training of deep learning models without the need for dedicated experimental setups and extensive data collection.
However, accurately simulating charge transport in silicon sensors remains a challenge.
For example, the spatial resolution achieved by deep learning models trained on simulated data often falls short of that attained by models trained on experimental data \cite{EM_deepLearning2023}. 
This discrepancy is primarily attributed to limitations in simulation accuracy, particularly the neglect of charge repulsion effects during transport.

Charge transport in silicon sensors, initiated by energy deposition from incident particles, forms the basis of detector signal and is described by the continuity equation for charge carriers:
\begin{equation}
\frac{\partial \rho}{\partial t} = D \Delta \rho - \nabla \cdot (\rho \mu \vec{E}) \label{eq1}
\end{equation}
where $\rho$ represents the charge carrier density, $D = \mu kT/e$ is the diffusion coefficient determined by the Einstein relation, $\mu$ is the charge carrier mobility, $k$ is the Boltzmann constant, $T$ is the tempreature, $e$ denotes the unit charge, and $\vec{E}$ denotes the electric field \footnote{
In conventional silicon detectors, recombination and generation of charge carriers are negligible and thus omitted from this discussion.}.
% By treating the drift of the charge cloud due to the applied bias voltage independently and neglecting the repulsion effect, 
% The charge carrier distribution can be approximated by a Gaussian profile with a standard deviation $\sigma = \sqrt{2Dt}$ if the repulsion effect is neglected.
% In a coordinate system centered on the drifting charge cloud due to the applied bias voltage, 
The charge cloud, consistting of charge carriers, drifts towards the electrodes driven by the applied bias voltage.
In a coordinate system centered on the drifting charge cloud, by neglecting the repulsion effect, Eq. \ref{eq1} reduces to $\frac{\partial \rho}{\partial t} = D \Delta \rho$.
The solution is a Gaussian with a standard deviation $\sigma = \sqrt{2Dt}$ in one dimension, where $t$ is the drift time.
% By treating the drift and diffusion of the charge carriers independently, the charge carrier distribution can be approximated by a Gaussian profile with a standard deviation $\sigma = \sqrt{2Dt}$, where $t$ is the drift time.
This approximation is widely adopted in existing simulation frameworks \cite{SPANNAGEL2018164,KDetSim}, and remains reasonably valid for minimum ionizing particles (MIPs) \footnote{See Section \ref{sec:Quantitative_repulsion} for details.}, where the resulting charge cloud is sufficiently spread out.
However, for X-rays with energies of more than a few keV, the point-like nature of energy depositions leads to highly concentrated charge clouds.
Under such conditions, the repulsion effect becomes significant and cannot be neglected, particularly within the first couple of nanoseconds \cite{1umCT, Medipix}.

Previous studies \cite{G4Medipix, s21041550} introduced a time dependent diffusion coefficient $D' = D + \frac{\mu N e}{24 \pi^{3/2}\epsilon_0 \epsilon_r \sigma(t)}$ \cite{BENOIT2009508} to account for the repulsion effect.
Here, the electric field dependency of the carrier mobility $\mu$ was ignored, and the results were primarily validated using a high-Z sensor with relatively large 225 \textmu m and 75 \textmu m pixel pitches.
Another previous study \cite{OldRepulsionStudy} employed a numerical method to simulate charge transport in silicon sensors with the repulsion effect included.
However, due to the substantial simulation time---in the order of several hours per simulation run---the simulation results were limited to only a few specific scenarios, preventing widespread adoption within the research community. 
In our earlier work \cite{SimuProceeding2024}, we introduced a time-stepping Monte Carlo simulation based on a simplified spherical method to effectively incorporate charge repulsion, and primarily validated it using X-ray fluorescence measurements.
Nevertheless, several limitations emerged, mainly due to the polychromatic nature of the X-ray fluorescence source.

In this work, we introduce a comprehensive time-stepping Monte Carlo simulation method that calculates charge repulsion among carriers using a brute-force manner accelerated by GPU computing.
The simplified spherical model from our previous work \cite{SimuProceeding2024} is briefly reviewed.
We improve the quality of experimental data by using a monochromatic X-ray beam collected at the METROLOGIE beamline \cite{METROLOGIE} of the SOLEIL synchrotron.
Careful calibrations and analysis procedures were applied to rigorously validate the simulation methods.
Furthermore, we refine the modeling and parameterization scheme to enable efficient generation of X-ray simulation events without compromising the simulation quality.
A systematic comparison between simulation and measurement of energy spectra of single pixels is conducted for both simulation methods across various configurations, including different sensor thicknesses, bias voltages, and photon energies.
Both methods achieved excellent agreement with measurements with a mean absolute deviation of less than 4\% for spectra ranging between 1 keV up to the beam energy, demonstrating the robustness and precision of the developed simulation techniques.
% This work will be integrated into the Allpix Squared framework \cite{SPANNAGEL2018164} as a dedicated charge propagation module.

This paper is structured as follows: Section 2 details the core algorithms of the two charge transport simulation methods.
In Section 3, we describe the modeling and parameterization scheme applied to the charge carrier distribution.
Section 4 outlines the measurement setup, data anaysis, and validation results.
In Section 5, we discuss the remaining discrepancy and the quantitative impact of the repulsion effect, and explore applications of these simulation methods, leading to the conclusions in Section 6.

\section{Charge transport simulation methods}\label{}
This section presents the core algorithms of two time-stepping Monte Carlo simulation methods. 
X-rays were selected as the incident particles for studying charge transport in silicon detectors due to their localized energy deposition and wide range of scientific applications.
Charge carriers are generated following the absorption of an X-ray photon, with a conversion coefficient of 3.62 eV per electron-hole pair.
We simulate the charge transport in p-in-n silicon sensors where the holes are collected by the readout electrodes, which is the case for the MÖNCH detector.
The sensor temperature is 34.7°C, as measured by a PT100 sensor mounted directly to a silicon sensor, while the chiller is operated at 20.0°C.
The time step is fixed at $\delta t = 10\ \text{ps}$ throughout the simulation for both methods, which is sufficiently small compared to the typical charge transport time of $O(10\ \text{ns})$ to ensure numerical accuracy.

\subsection{Comprehensive simulation method}
Both electrons and holes are simulated as individual charge carriers.
The initial spatial distribution is approximated as a Gaussian, with a standard deviation given by the semi-empirical relation \cite{Rele, ReleToSigma}:
\begin{equation}
    \sigma = R_{e^-}/\sqrt{15} = 0.0044 \cdot E_{\rm deposit}^{1.75}
\end{equation}
where $R_{e^-}$ is the Bethe range of the initial photoelectron.
This expression is valid within the energy range of 5-25 keV.
At each time step, the $x$-coordinate of the $i$-th charge carrier is updated as:
\begin{equation}
    \label{eq:comprehensive_x}
    x_{i, t+\delta t} = x_{i,t} \pm \sqrt{2D\delta t} + \mu\vec{E}_{\text{repulsion},i} \cdot \vec{e}_x\delta t
\end{equation}
In this equation, the second term describes diffusion, modeled as a random walk, while the third term accounts for the Coulomb repulsion\footnote{The Coulomb atrraction between opposite types of charge carriers is also included. However, this effect rapidly becomes negligible as the separation between two charge clouds increases.}.
The repulsion electric field is calculated as:
\begin{equation}
    \vec{E}_{\text{repulsion}, i} = \Sigma_{j\neq i} \frac{q_i q_j}{4\pi\epsilon_0 \epsilon_r r_{ij}^3} \vec{r}_{ij} 
\end{equation}
Here, $q=\pm e$ denote the charge of the carrier, $\epsilon_0$ is the vacuum permittivity, $\epsilon_r=11.7$ is the relative permittivity of silicon, and $\vec{r}_{ij}$ is the vector from carrier $j$ to carrier $i$.
The $y$-coordinate is updated in the same manner.
For the $z$ coordinate, the update further includes the drift component:
\begin{equation}
    \label{eq:comprehensive_z}
    z_{i, t+\delta t} = z_{i,t} \pm \sqrt{2D\delta t} + \mu\cdot(\vec{E}_{\text{drift},i,} + \vec{E}_{\text{repulsion},i}) \cdot \vec{e}_z\delta t
\end{equation}
The drift electric field $\vec{E}_{\text{drift}}$ is approximated as:
\begin{equation}
    \vec{E}_{\text{drift},i} = (\frac{V_{\rm bias} - V_{\rm depletion}}{H} + \frac{2 V_{\rm depletion}}{H} \cdot \frac{z_i}{H}) \cdot \vec{e}_z
\end{equation}
where $V_{\rm bias}$ is the applied bias voltage, $V_{\rm depletion}$ is the depletion voltage, and $H$ is the sensor thickness, assuming a uniform bulk doping concentration in the silicon sensor.
This approximation is well-suited for hybrid pixel detectors with thick silicon substrates.
The Jacoboni-Canali model \cite{Jacoboni-Canali}, which depends on the magnitude of the total electric field $|\vec{E}_{\rm drift} + \vec{E}_{\rm repulsion}|$ and temperature, is used to model the mobility of holes and electrons. 
The simulation proceeds until all holes reach the electrodes.
The complete algorithm is presented in Appendix \ref{sec:appendix_algorithm}.

The computational complexity is $O(N^2)$, where $N$ is the number of charge carriers, due to the pairwise calculation of the repulsion electric field.
This computational cost is mitigated by GPU acceleration.
Using a PyTorch-based \cite{PyTorch} implementation, simulating charge transport from one single photon typically takes less than one minute on an NVIDIA RTX 4090 GPU, which is around 25 times faster than running on a 16-core CPU.

\subsection{Simplified spherical model}
The simplified spherical model, introduced in our previous work \cite{SimuProceeding2024}, is briefly summarized here.
This approach assumes spherical symmetry for the charge cloud, which simplifies the calculation of the repulsion electric field at a radial distance $r$ from the cloud center, expressed as:
\begin{equation}
    \vec{E}_{\text{repulsion}}(r) = \frac{\Sigma_{r_i<r} q_i}{4\pi\epsilon_0 \epsilon_r r^3} \vec{r}
\end{equation}
In this model, only the hole coordinates are updated, following the same formulations as described in Eq. \ref{eq:comprehensive_x} and Eq. \ref{eq:comprehensive_z}.
This symmetry-based simplification significantly reduces the computational complexity from $O(N^2)$ to $O(N)$, reducing typical simulation time to a few seconds when executed on a single CPU core.

\section{Modeling and parameterization}\label{}
Despite the computational efficiency of the simplified spherical model, the simulation time remains considerable when generating a large number of events.
To efficiently produce X-ray simulation events, accurate modeling and parameterization of charge transport behavior across various absorption depths are essential.
In prior work \cite{SimuProceeding2024}, we effectively modeled the charge transport behavior using Generalized Gaussian Distribution (GGD).
Simulation events were generated at discrete absorption depths based on the precomputed GGD parameters.
In this study, to enable sample generation at arbitrary absorption depths, we refine the scheme by parametrizing the modeling parameters as continuous functions of the total drift time.
This section illustrates the modeling and parameterization procedure using results from the comprehensive simulation method for 25~keV photon energy in a $H=320$ \textmu m sensor thickness biased at 150 V with a depletion voltage of 29.6~V.

\subsection{Modeling}
The distribution of collected holes along $x$-direction is modeled using a GGD:
\begin{equation}
    f(x) = \frac{\beta}{\alpha \Gamma(1/\beta)} \text{exp}(-(x/\alpha)^\beta)
\end{equation}
where $\alpha$ is the scale parameter, $\beta$ is the shape parameter, and $\Gamma$ denotes the gamma function.
When $\beta=2$, the GGD reduces to a Gaussian distribution.
A value of $\beta>2$ indicates a broader peak compared to Gaussian, characteristic of the repulsion effect in our case.
Due to the symmetry, the $y$-coordinate distribution shares the same GGD parameters.

\begin{figure}
\centering
\includegraphics[width=0.45\textwidth]{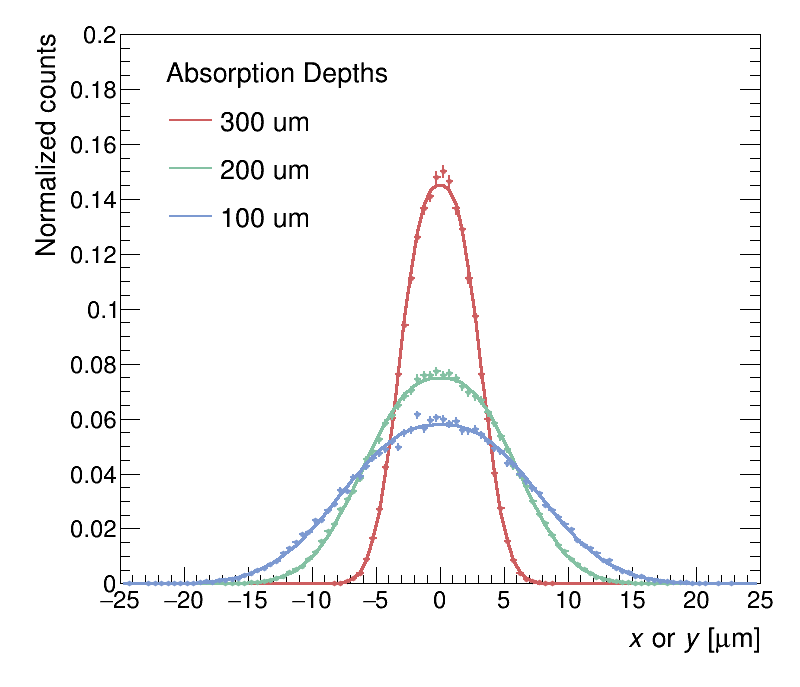}
\caption{Hole distributions at the end of the simulation for absorption depths of 300, 200, and 100 \textmu m.
Both $x$-and $y$-coordinates from ten independent simulation runs are combined into the histogram for statistics.
The error bars represent the statistical uncertainties.
The Generalized Gaussian Distribution (GGD) fits are shown in solid lines.
\label{fig:GGD}
}
\end{figure}
Fig. \ref{fig:GGD} shows hole distributions at absorption depths of 300, 200, and 100 \textmu m, along with GGD fits in solid lines.
To reduce statistical fluctuations, both $x$ and $y$ coordinates from ten independent simulation runs were combined into the histogram.
The fitted value of $\beta=2.63$, 2.47, 2.40 reflect the impact of the repulsion effect.
The resulting chi-squared per degree of freedom ($\chi^2/\text{NDF}$) of less than 2 for all tested cases confirm the adequacy of the GGD model.

\subsection{Parameterization} 
The GGD parameters $\alpha,\beta$ were parameterized as functions of the approximated drift time $t$, following the approach employed by the $ProjectionPropagation$ module in Allpix Squared \cite{SPANNAGEL2018164}:
\begin{equation}
    \label{eq:drift_time}
    t = \int \frac{1}{v}dz \approx \frac{1}{\mu_0} [\frac{\ln(E_{\text{drift}}(z))}{k} + \frac{z}{E_c}]\lvert ^H_h
\end{equation}
$\mu_0$ and $E_c$ are the parameters in the Jacoboni-Canali model; $k$ is the derivative of the drift electric field; $h$ is the photon absorption depth, and $H$ is the sensor thickness.
\begin{figure}
    \centering
    \includegraphics[width=0.45\textwidth]{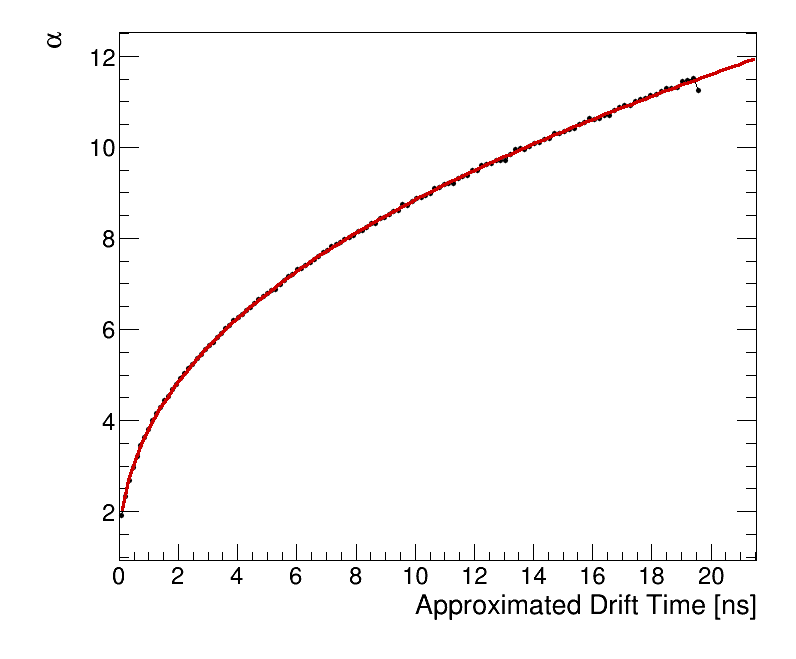}
    \includegraphics[width=0.45\textwidth]{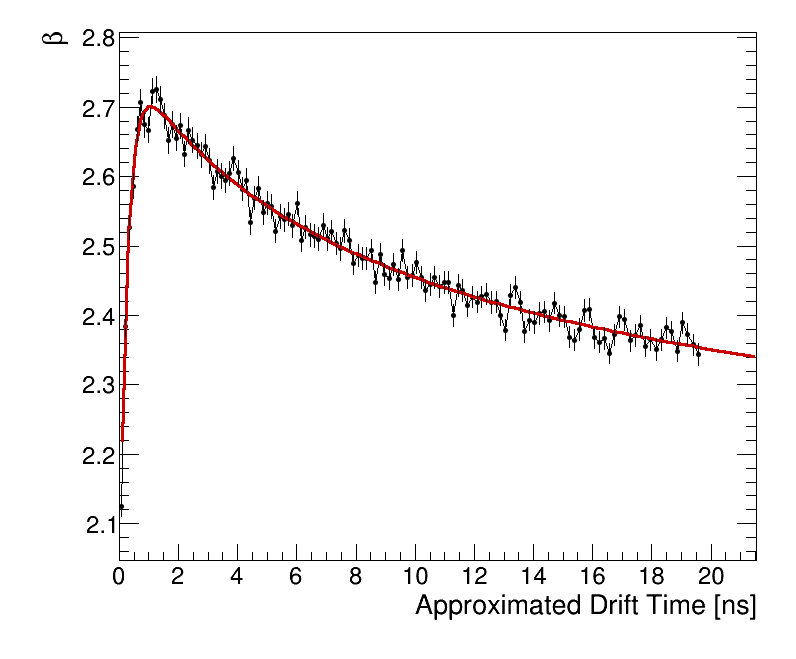}
    \caption{GGD parameters as functions of the approximated drift time $t$. (Top) Scale parameter $\alpha(t)$; (bottom) shape parameter $\beta(t)$. The fitted functions, given by Eq. \ref{eq:alpha_over_t} and Eq. \ref{eq:beta_over_t}, are shown in red. 
    \label{fig:GGD_param}
    }
\end{figure}
The GGD parameters $\alpha$ and $\beta$ are then plotted as functions of the approximated drift time $t$ in Fig. \ref{fig:GGD_param}, with uncertainties derived from the fitting.
A sharp increase in $\beta$ at short drift times reflects the strong repulsion effect when charge carriers are densely packed at the beginning of the transport.
% As the repulsion effect weakens, the $\beta$ parameter gradually approaches 2, i.e., the hole distribution is converging to a Gaussian distribution dominated by the diffusion.
As the repulsion effect diminishes with increasing spatial separation, the $\beta$ value gradually approaches 2, indicating convergence toward a diffusion-dominated Gaussian distribution.

The following functions were proposed to fit $\alpha(t)$ and $\beta(t)$ as functions of the approximated drift time $t$:
\begin{equation}
    \label{eq:alpha_over_t}
    \alpha(t) = \text{p}_{\alpha,0} + \text{p}_{\alpha,1} \sqrt{t} + \text{p}_{\alpha,2} t + \text{p}_{\alpha,3} t^2
\end{equation}
and
\begin{equation}
    \label{eq:beta_over_t}
    \beta(t) = 2 + \text{p}_{\beta,0}(t-p_{\beta,1})^{\text{p}_{\beta,2}} + \text{p}_{\beta,3}\text{exp}(\text{p}_{\beta,4}t)
\end{equation}
where $\text{p}_{\alpha,0}$ to $\text{p}_{\alpha,3}$ and $\text{p}_{\beta,0}$ to $\text{p}_{\beta,4}$ are the nine free parameters determined by fitting.
Eq. \ref{eq:beta_over_t} is designed to ensure that $\beta(t)$ starts from 2 at $t=0$ and approaches 2 as $t$ increases, reflecting the diminishing repulsion effect.
Both parameterizations yielded satisfactory fits with $\chi^2/\text{NDF}$ of 2.0 and 1.0, respectively.
This scheme effectively captures the charge transport behavior across the sensor depth using only nine parameters. 
It is important to note that these parameters are implicitly dependent on experimental factors such as photon energy, sensor thickness, depletion voltage, and bias voltage.

\section{Measurement and validation}\label{}
Previous validation results based on Cu and Ag X-ray fluorescence measurements \cite{SimuProceeding2024} were limited by the polychromatic nature of the X-ray fluorescence source and the undefined portions of $K\alpha_{1}$, $K\alpha_{2}$, and $K\beta_{1}$ X-ray photons.
In this study, we improved the measurement quality by collecting monochromatic X-ray photons using a synchrotron radiation source.
Energy spectra of single pixels were obtained under various experimental conditions, including sensor thickness, bias voltage, and photon energy.
Corresponding simulation datasets were generated using parameters derived from parametrization procedure described above, applying both the comprehensive and simplified simulation methods.
Identical analysis procedures were applied to both simulation and experimental datasets, enabling rigorous and systematic comparisons.

\subsection{Measurement and analysis}
We conducted measurements using a monochromatic, low flux X-ray beam incident perpendicularly on the sensor surface at the METROLOGIE beamline \cite{METROLOGIE} of the SOLEIL synchrotron.
The beam spot size exceeded 1~$\text{cm}^2$, ensuring flat-field illumination over the sensor's region of interest.
Silicon sensors with thicknesses of 320~\textmu m and 650~\textmu m, bump-bonded to MÖNCH03 readout chips \cite{MONCH}, were used for photon collection.
The corresponding depletion voltages are 29.6 and 34.7 V, respectively.
The 650~\textmu m sensor was biased at 150 V bias voltage, while the 320~\textmu m sensor was measured at bias voltages of 90 V and 150 V.

Pixel-wise gain and non-linearity calibrations were performed utilizing the backside-pulsing method \cite{Mezza_2016}.
Clusters were defined with the local maximum pixel signal at the center.
Events containing dead or noisy pixels within the 5x5 pixel cluster were discarded.
Pile-up events were excluded by requiring that no pixel in the annular region between 3x3 and 7x7 pixel clusters had a readout exceeding three times its noise level.
Additionally, a $\pm$1 keV energy selection window around the beam energy was applied.
The resulting energy spectra of single pixels were composed of pixels from over one million 3x3 pixel clusters.

\begin{figure*}
    \centering
    \includegraphics[width=1\textwidth]{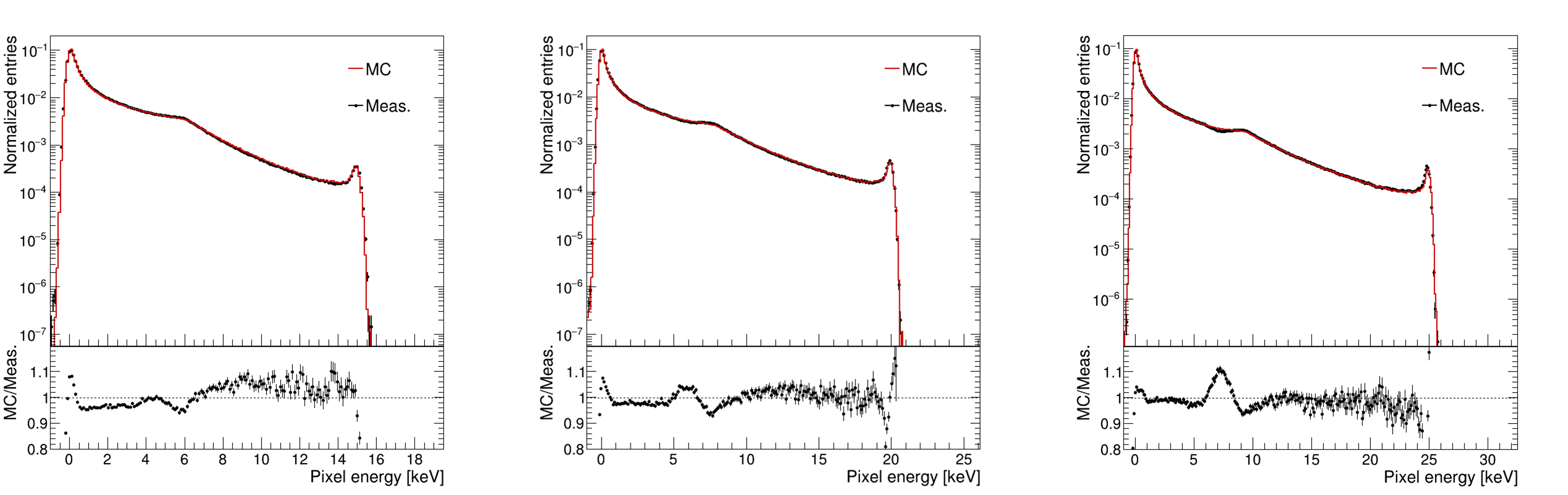}
    \caption{
        Energy spectra of single pixels from the comprehensive simulation (red) and measurements (black) of 650 \textmu m thick sensor at 150 V bias voltage, for (left) 15 keV, (middle) 20 keV, and (right) 25 keV X-ray photons.
        The bin width is 0.1 keV, and the error bars represent the statistical uncertainties.
        The bottom panel displays the corresponding ratio of simulation to measurement.
    }
    \label{fig:validation_650_comprehensive}
\end{figure*}
\subsection{Simulation generation and validation results}
For each experimental configuration---defined by photon energy, bias voltage, and sensor thickness---128$\times$10 charge transport simulations were performed, corresponding to 128 discrete photon absorption depths, each repeated ten times for statistics.
Each simulation set required several hours to complete using eight NVIDIA A100 GPUs when employing the comprehensive simulation method.
In contrast, simulations using the simplified spherical model were completed in about 2 minutes using a 16-core CPU.
Following simulation, modeling and parameterization procedures described in Section 3 were conducted to extract the parameters in Eq. \ref{eq:alpha_over_t} and Eq. \ref{eq:beta_over_t} for each configuration.

Subsequently, one million single X-ray events were generated for each configuration by reversing the parameterization procedure.
Photon absorption depths $h$ were sampled from an exponential distribution, and the approximated drift times $t$ were calculated using Eq. \ref{eq:drift_time}.
A Fano factor of 0.13 was applied to model the statistical fluctuations in the number of charge carriers.
Charge carrier distributions were sampled based on the derived $\alpha(t)$ and $\beta(t)$.
Random sub-pixel shifts were introduced in the $x$ and $y$ coordinates to reproduce the flat-field illumination.
By converting the carrier number within each pixel to energy and incorporating realistic detector noise of around 0.13 keV, the detector energy readouts were obtained.
Identical selection criteria and energy cuts described above were applied to the simulation events to derive the energy spectra of single pixels.
Generating one million events takes approximately 2 minutes using a 16-core CPU.

\begin{figure*}
    \centering
    \includegraphics[width=1\textwidth]{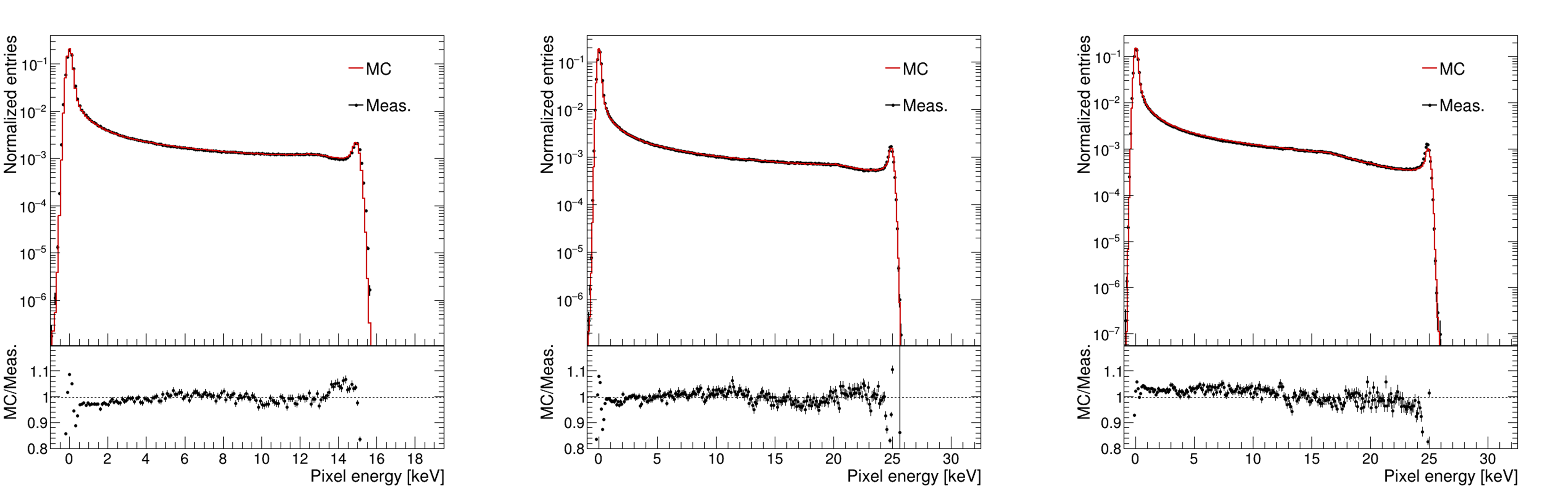}
    \caption{
        Energy spectra of single pixels from the comprehensive simulation (red) and measurements (black) for 320 \textmu m thick sensor at 150 V bias voltage, for (left) 15 keV, (middle) 25 keV X-ray photons, and at 90 V bias voltage for (right) 25 keV X-ray photons.
        The bin width is 0.1 keV, and the error bars represent the statistical uncertainties.
        The bottom panel displays the corresponding ratio of simulation to measurement.
    }
    \label{fig:validation_320_comprehensive}
    \centering
    \includegraphics[width=1\textwidth]{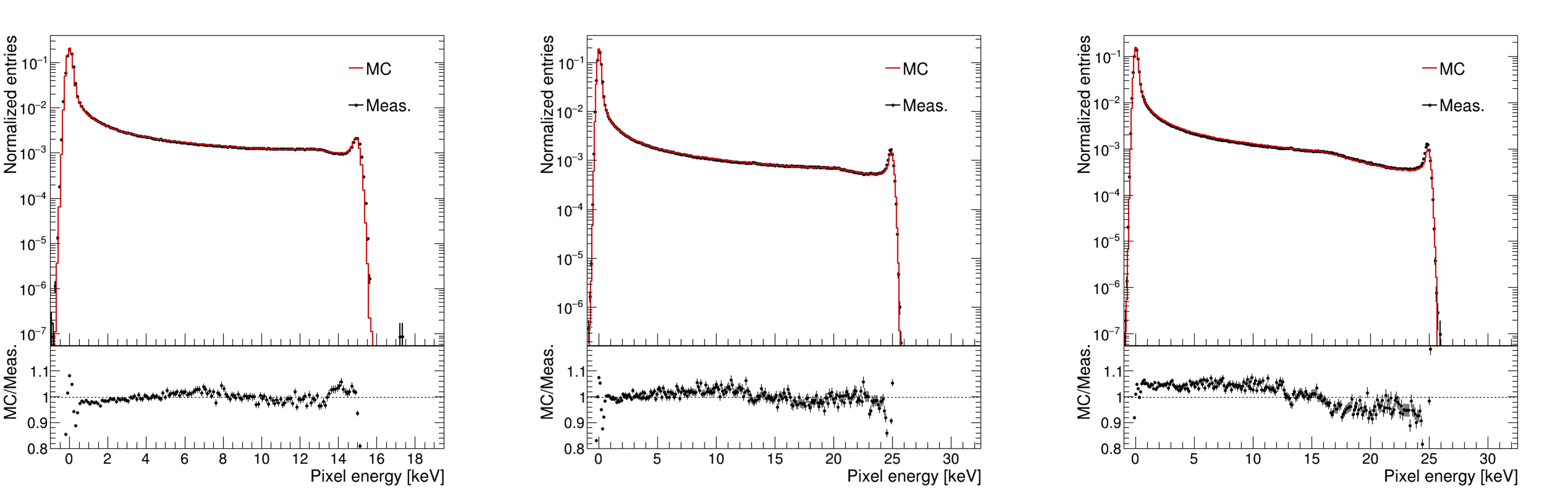}
    \caption{
        Energy spectra of single pixels from the simplified simulation (red) and measurements (black) for 320 \textmu m thick sensor at 150 V bias voltage, for (left) 15 keV, (middle) 25 keV X-ray photons, and at 90 V bias voltage for (right) 25 keV X-ray photons.
        The bin width is 0.1 keV, and the error bars represent the statistical uncertainties.
        The bottom panel displays the corresponding ratio of simulation to measurement.
    }
    \label{fig:validation_320_simplified}
\end{figure*}

Validation results obtained with the 650 \textmu m sensor at 150 V bias voltage, using 15 keV, 20 keV, and 25 keV X-ray photons, are presented in Fig. \ref{fig:validation_650_comprehensive}.
For all tested configurations, the ratio of the simulated to experimental energy spectra centers remained close to 100\% typically within a $\pm$10\% margin, demonstrating satisfactory agreement.
Specifically, the mean absolute deviations between simulation and measurement from 1 keV up to the beam energy are less than 4\% and the $\chi^2/\text{NDF}$ values were under 25.
For the 320 \textmu m sensor, as shown in Fig. \ref{fig:validation_320_comprehensive}, similarly high consistency is obtained for 15 keV and 25 keV X-ray photons at 150 V bias voltage, and for 25 keV X-ray photons at 90 V bias voltage, with the mean absolute deviations remaining below 3\%, with $\chi^2/\text{NDF}$ values under 20.
A comparison using the simplified spherical model is shown in Fig. \ref{fig:validation_320_simplified}.
This model demonstrates comparable consistency but with significantly reduced simulation time, highlighting its efficiency and practical utility.

\section{Discussion and conclusion}\label{}

\subsection{Quantitative understanding of the repulsion effect}
\label{sec:Quantitative_repulsion}
\begin{figure}
    \centering
    \includegraphics[width=0.45\textwidth]{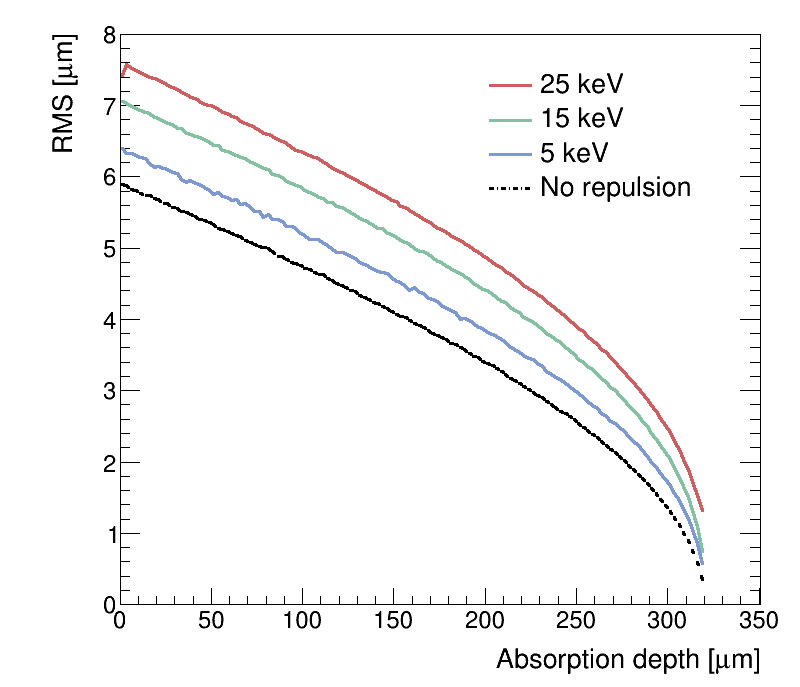}
    \caption{
        Root mean square (RMS) of hole distributions after transport simulation as a function of the absorption depth for a 320 \textmu m sensor at 150 V bias voltage, shown for different photon energies.
        The dashed line represents the limiting case where the repulsion effect is removed and the initial Gaussian distribution width $\sigma$ is set to zero, approximating the behavior at 0 keV photon energy.
    }
    \label{fig:RMS_depth}
\end{figure}

We quantitatively investigated the influence of the repulsion effect on charge transport using the comprehensive simulation method applied to a 320 µm-thick sensor biased at 150 V, a commonly employed configuration for hybrid pixel detectors.
Fig. \ref{fig:RMS_depth} illustrates the root mean square (RMS) of the collected hole distribution as a function of photon absorption depth for different photon energies.
To examine the theoretical limiting case of zero photon energy (0 keV), we conducted additional simulations with a zero standard deviation of the initial Gaussian distribution and the repulsion effect removed.
This scenario is indicated in Fig. \ref{fig:RMS_depth} by a dashed line for reference.
In the shallow absorption region of the plot, the repulsion effect rapidly broadens the charge cloud, with the extent of broadening strongly dependent on photon energy.
At longer drift distances, where the repulsion effect diminishes toward the end of the transport, the differences in RMS among different photon energies become stable.

For a more quantitative understanding, the RMS values in Fig. \ref{fig:RMS_depth} were weighted by the absorption probability and summarized in Table \ref{tab:RMS}.
To provide a direct comparison, we additionally computed RMS values from simulations performed without the repulsion effect.
For photon energies as low as 5 keV, the repulsion effect increases the overall RMS of the charge carrier distribution by approximately 9\% relative to simulations without repulsion.
At higher photon energies,  the weighted RMSs are smaller due to the increased attenuation length; however, the relative impact of the repulsion effect becomes more pronounced, as expected.
This phenomenon is consistent with the behavior observed in Fig.~\ref{fig:GGD_param}b.

\begin{table}
    \centering
    \caption{Root mean square (RMS) of holes after transport simulation of 320 \textmu m sensor at 150 V bias voltage for different photon energies.
    The RMS is weighted by the attenuation probability at different absorption depths.}
    \begin{tabular*}{\tblwidth}{@{} LLLL@{} }
        \toprule
        Weighted RMS & 5 keV & 15 keV & 25 keV\\
        \midrule
        Without repulsion & 5.7 \textmu m & 4.3 \textmu m & 4.3 \textmu m \\
        With repulsion & 6.2 \textmu m & 5.3 \textmu m & 5.6 \textmu m \\
        \bottomrule
    \end{tabular*}
    \label{tab:RMS}
\end{table}

The repulsion effect on minimized Ionizing Particles (MIPs) was also primarily tested using the comprehensive simulation method.
We used a 50 \textmu m-thick sensor with a 20 V depletion voltage, operated at a bias voltage of 50 V.
For a quick validation, we kept the linear approximation for the drift electric field to emulate the fully depleted monolithic active pixel sensors (MAPS).
Assuming a MIP trajectory perpendicular to the sensor surface, the initial charge carrier distribution was modeled as a uniform with a density of 80 pairs per \textmu m across the sensor thickness.
In the $x$ and $y$ directions, the initial distribution followed a Gaussian profile with a standard deviation of 1 nm.
The RMS of the collected electron distribution was 1.27 \textmu m with the repulsion effect included, 2.4\% larger than the 1.24 \textmu m RMS obtained when repulsion was neglected.
This difference becomes even less significant when the incident MIP trajectory is not perpendicular to the sensor surface or when the MAPS sensor is not fully depleted.
These results support the conventional assumption that repulsion effects can be neglected for MIPs, and that the Gaussian approximation remains valid.

\subsection{Limitations of the simulation methods}
Comparing the results in Fig. \ref{fig:validation_320_simplified} with our previous work \cite{SimuProceeding2024}, we observed substantial improvements in the agreement between simulations and measurements, primarily due to the use of monochromatic X-ray beam.
Additionally, the comprehensive simulation method eliminates other previously identified sources of discrepancies, such as the spherical symmetry assumption and the boundary condition.
However, the comparable performance of the simplified spherical model suggests that the dominant remaining source of discrepancy is the intrinsic uncertainty—up to 10\%—in the Jacoboni-Canali mobility model \cite{Jacoboni-Canali}.
% Other potential sources of discrepancy include parameter uncertainties, such as the 3.62 eV per carrier pair conversion coefficient and the relative permittivity of silicon (11.7), as well as uncertainties in sensor thickness and deviations from ideal perpendicular X-ray incidence.
Other potential sources of discrepancy include non-uniformities of the electric field distribution due to the pixel structure as well as the non-uniform doping concentration in the sensor bulk.

It is important to note that the semi-empirical Gaussian model employed for the initial charge carrier distribution has been validated within the photon energy range of 5-25 keV, which adequately covers typical operational conditions for hybrid pixel detectors.
For photon energies exceeding this range, the extended Bethe range of photoelectrons leads to broader charge cloud spreads, which are not well described by a simple Gaussian distribution.

% The two charge transport simulation methods, along with the refined parameterization scheme, enable efficient and accurate generation of X-ray simulation events, e.g., for interpolation and deep learning studies using hybrid pixel detectors.
\subsection{Potential applications of the simulation methods}
The two charge transport simulation methods, along with the refined parameterization scheme, enable efficient and accurate generation of X-ray simulation events.
In X-ray interpolation tasks, the interpolation mapping can be derived from simulation events, eliminating the need for time-consuming flat-field data acquisition.
Moreover, simulation inherently provides access to ground truth information, which is challenging or even impossible to obtain experimentally.
For example, in deep learning-based position reconstruction tasks \cite{EM_deepLearning2023}, labeling experimental training samples with the true incident position required substantial effort and relied on the specific alignment of the electron microscope beam.
In contract, accurate simulation events with access to ground truth information can significantly facilitate deep learning studies.

With appropriate extensions, the two simulation methods also have broad applicability across a range of detector technologies.
The comprehensive simulation method, which tracks individual charge carriers, can be adapted for detectors with complex geometries and electric field distributions, such as monolithic active pixel sensors (MAPS) and 3D sensor designs.
Incorporating the weighting field would further enable this approach to simulate transient signal responses.
Additionally, this method holds the potential for studying signal formation in high-Z sensors and low-gain avalanche sensors (LGAD), where additional processes such as charge trapping, detrapping, and multiplication processes must be considered.

On the other hand, the simplified spherical model, which assumes spherical symmetry of the charge cloud, achieves  comparable accuracy with significantly reduced computational cost in the case of silicon hybrid pixel detectors.
This method can be extended for the more detectors where the spherical symmetry of charge cloud is also generally valid, such as microstrip detectors.
% With certain computational pwoer, such as a powerful cluster, it is feasible to directly generate simulation events of arbitrary energies and absorption depths.
Provided adequate computational resources (e.g., a high-performance computing cluster) this model enables the generation of simulation events across arbitrary photon energies and absorption depths, without relying on precomputed parameters for a specific energy and depth.
% which benefits the simulation for electrons of energies higher than 50 keV where the energy deposited along the trajectory varies significantly.
This capability is especially valuable in simulating high-energy electrons (e.g., >50 keV), whose energy deposition varies significantly along their penetration trajectories.

\section{Conclusion}
We have developed and validated two time-stepping Monte Carlo simulation methods that explicitly incorporate the repulsion effect among charge carriers in the charge transport. 
Both the comprehensive simulation approach, employing brute-force pairwise calculations accelerated by GPU computing, and the computationally efficient simplified spherical model were rigorously validated across various configurations, including different sensor thicknesses, bias voltages, and photon energies.
Additionally, we introduced a refined modeling and parameterization scheme that enables efficient generation of X-ray simulation events based on these methods.
% Systematic validation using monochromatic X-ray data demonstrated excellent agreement between simulated and measured pixel-energy spectra across various configurations, including different sensor thicknesses, bias voltages, and photon energies.

These two simulation approaches represent powerful tools for investigating charge transport properties, quantifying the repulsion effect, and guiding the optimization of detector designs.
Coupled with the parameterization scheme, they facilitate efficient and accurate event generation, beneficial for interpolation tasks and deep learning studies. 
Integration of these methods into the widely used Allpix Squared simulation framework is planned, offering the research community a robust, accurate, and convenient simulation tool to support detector developments.

\section*{Research data}
The simulation source codes, measured energy spectra of single pixels, and example simulation results are available at: \url{https://github.com/slsdetectorgroup/ChargeTransportSimulation}.

% \section*{Declaration of Competing Interest}
% The authors declare that they have no known competing financial interests or personal relationships that could have appeared to influence the work reported in this paper.

\section*{Acknowledgments}
The authors gratefully acknowledge Marie Andrae and Arkadiusz Dawiec for their support during the beamtime, as well as the assistance provided by the METROLOGIE beamline staff and the detector group at the SOLEIL synchrotron.
We also thank Håkan Wennlöf and Simon Spannagel for the valuable discussions regarding the development and integration of the simulation methods into the Allpix Squared framework.

This manuscript was prepared with the assistance of ChatGPT for language editing and grammatical refinement.

% \bibliographystyle{model1a-num-names}
% \bibliographystyle{cas-model2-names}
\bibliographystyle{elsarticle-num}
% \bibliographystyle{unsrt}

\bibliography{cas-refs}



\appendix
\section{Appendix: algorithm of the comprehensive simulation method}
\label{sec:appendix_algorithm}

\begin{algorithm*}
    \caption{Comprehensive Monte Carlo Charge Transport Simulation}
    \begin{algorithmic}[1]
    \State \textbf{Input:} Photon energy $E$, absorption depth $h$, sensor thickness $H$, sensor temperature $T$, bias voltage $V_{\text{bias}}$, depletion voltage $V_{\text{depletion}}$
    \State \textbf{Output:} Final positions of charge carriers
    \Statex
    
    \State Initialize $N = \text{round}(E / 3.62)$ charge carriers with 3D Gaussian distribution ($\sigma_{\text{1D}} = 0.0044 \cdot E^{1.75}$)
    
    \While{not all holes reach readout electrode}
        \State Compute repulsion electric field for each carrier pair: $\vec{E}_{\text{repulsion}, i \ne j} = \frac{q_i q_j}{4\pi \varepsilon_0 \varepsilon_r r_{ij}^3} \vec{r}_{ij}$
        \For{each carrier $i = 1$ to $N$}
            \State Compute repulsion field: $\vec{E}_{\text{repulsion}, i} = \sum_{j \ne i} \frac{q_i q_j}{4\pi \varepsilon_0 \varepsilon_r r_{ij}^3} \vec{r}_{ij}$
            \State Compute drift field: $\vec{E}_{\text{drift}, i} = \left( \frac{V_{\text{bias}} - V_{\text{depletion}}}{H} + \frac{2V_{\text{depletion}}}{H^2} z_i \right)\vec{e}_z$
            \State Compute total field: $\vec{E}_{\text{total}, i} = \vec{E}_{\text{drift}, i} + \vec{E}_{\text{repulsion}, i}$
            \State Compute mobility $\mu$ using Jacoboni–Canali model based on $|\vec{E}_{\text{total}, i}|$ and temperature $T$
            \State Update position:
            \begin{align*}
                x_i &\gets x_i \pm \sqrt{2D\delta t} + \mu \vec{E}_{\text{repulsion}, i} \cdot \vec{e}_x \delta t \\
                y_i &\gets y_i \pm \sqrt{2D\delta t}+ \mu \vec{E}_{\text{repulsion}, i} \cdot \vec{e}_y \delta t \\
                z_i &\gets z_i \pm \sqrt{2D\delta t} + \mu \vec{E}_{\text{total}, i} \cdot \vec{e}_z \delta t
            \end{align*}
        \EndFor
    \EndWhile
    
    \State \Return Final positions of all holes
    \end{algorithmic}
\end{algorithm*}

\end{document}